%% file: QC_AED.tex
\documentclass[conference]{IEEEtran}

\renewcommand\IEEEkeywordsname{Index Terms}
\usepackage{graphicx}
\usepackage{subcaption}
\usepackage{tikz,xparse}

\usetikzlibrary{patterns}

\usetikzlibrary{dsp,chains}
\usetikzlibrary{matrix}
\usepackage{mathptmx}
\usepackage{verbatim}
\usepackage{calc}
\usepackage{ifthen}
\usepackage{xifthen}
\usepackage{cancel}
\usepackage{bm}
\usepackage{verbatim}
\usepackage{multirow}
\usepackage{cite}
\usepackage[hyphens]{url}
\usepackage[nolist]{acronym} 
\usepackage{pgfplots}
\usetikzlibrary{arrows,shapes,graphs,graphs.standard,quotes,arrows.meta,decorations.markings,backgrounds}
\pgfplotsset{compat=newest}
\usepackage[bookmarks=false]{hyperref}
\usepackage{units}
\usepackage{amsmath, amsbsy, amssymb, latexsym }
\hypersetup{bookmarksdepth=-2}
\usepackage{comment}
\usepackage[utf8]{inputenc}
\usepackage{xcolor}
\usepackage{enumitem}
\usepackage[normalem]{ulem}

\usepackage{marginnote}
\tikzset{>=latex}

\input{corporateColours.tex}
\input{nodesAndStyles.tex}

\newcommand\new[1]{#1}
\IEEEoverridecommandlockouts

\input{macros}		

\begin{document}

\begin{NoHyper}
\title{Automorphism Ensemble Decoding of Quasi-Cyclic LDPC Codes by Breaking Graph Symmetries
}

\author{\IEEEauthorblockN{Marvin Geiselhart, Moustafa Ebada, Ahmed Elkelesh, Jannis Clausius and Stephan ten Brink }
\thanks{The authors are with the Institute of Telecommunications, Pfaffenwaldring 47, University of  Stuttgart, 70569 Stuttgart, Germany (e-mail: \{geiselhart,ebada,elkelesh,clausius,tenbrink\}@inue.uni-stuttgart.de)

This work is supported by the German Federal Ministry of Education and Research (BMBF) within the project Open6GHub (grant no. 16KISK019).
}
\vspace{-0.5cm}
}

\maketitle

\begin{acronym}
\acro{ML}{maximum likelihood}
\acro{BP}{belief propagation}
\acro{BPL}{belief propagation list}
\acro{LDPC}{low-density parity-check}
\acro{BER}{bit error rate}
\acro{SNR}{signal-to-noise-ratio}
\acro{BPSK}{binary phase shift keying}
\acro{AWGN}{additive white Gaussian noise}
\acro{BI-AWGN}{binary-input additive white Gaussian noise}
\acro{LLR}{log-likelihood ratio}
\acro{MAP}{maximum a posteriori}
\acro{BLER}{block error rate}
\acro{SCL}{successive cancellation list}
\acro{SC}{successive cancellation}
\acro{BI-DMC}{Binary Input Discrete Memoryless Channel}
\acro{CRC}{cyclic redundancy check}
\acro{CA-SCL}{CRC-aided successive cancellation list}
\acro{BEC}{Binary Erasure Channel}
\acro{BSC}{Binary Symmetric Channel}
\acro{BCH}{Bose-Chaudhuri-Hocquenghem}
\acro{RM}{Reed--Muller}
\acro{RS}{Reed-Solomon}
\acro{SISO}{soft-in/soft-out}
\acro{3GPP}{3rd Generation Partnership Project }
\acro{eMBB}{enhanced Mobile Broadband}
\acro{CN}{check node}
\acro{VN}{variable node}
\acro{GenAlg}{Genetic Algorithm}
\acro{CSI}{Channel State Information}
\acro{OSD}{ordered statistic decoding}
\acro{MWPC-BP}{minimum-weight parity-check BP}
\acro{FFG}{Forney-style factor graph}
\acro{MBBP}{multiple-bases belief propagation}
\acro{URLLC}{ultra-reliable low-latency communications}
\acro{DMC}{discrete memoryless channel}
\acro{SGD}{stochastic gradient descent}
\acro{QC}{quasi-cyclic}
\acro{AED}{automorphism ensemble decoding}
\acro{SPA}{sum-product algorithm}
\acro{SBP}{saturated BP}
\end{acronym}

\begin{abstract}
We consider \ac{AED} of \ac{QC} \ac{LDPC} codes. Belief propagation (BP)\acused{BP} decoding on the conventional factor graph is equivariant to the quasi-cyclic automorphisms and therefore prevents gains by AED. However, by applying small modifications to the parity-check matrix at the receiver side, we can break the symmetry without changing the code at the transmitter. This way, we can leverage a gain in error-correcting performance using an ensemble of identical \ac{BP} decoders, without increasing the worst-case decoding latency. The proposed method is demonstrated using \ac{LDPC} codes from the CCSDS, 802.11n and 5G standards and produces gains of 0.2 to 0.3 dB over conventional \ac{BP} decoding. \new{Compared to the similarly performing \ac{SBP}, the proposed algorithm reduces the average decoding latency by more than eight times.}
\end{abstract}
\acresetall

\section{Introduction}

Quasi-cyclic (QC)\acused{QC} \ac{LDPC} codes are the error-correction workhorse of modern communication systems (e.g., CCSDS, Wi-Fi 802.11n and 5G \cite{5GLDPC} standards), motivated by the presence of a well-understood, low-complexity \ac{BP} decoder.
Long \ac{LDPC} codes constructed using classical information theoretic design tools can closely approach the Shannon limit under \ac{BP} decoding \cite{CapLDPC}. 
However, in the short block-length regime (block-lengths of few hundreds of bits) \ac{LDPC} codes perform poorly when compared to other structured algebraic coding (e.g., \ac{BCH}, \ac{RM} and \ac{CRC}-aided polar codes), see~\cite{liva2016} for an exhaustive comparison.
The degraded error-rate performance can be attributed to the non-optimal \ac{BP} decoding algorithm (when compared to the \ac{ML} decoder) and the sub-optimality of the short length \ac{LDPC} code design.
The problem of designing short length \ac{LDPC} codes is out of the scope of this paper. 

In this paper, we are interested in enhancing the decoding algorithm itself without changing the code structure.
\new{We show ways of enhancing the error-rate performance under iterative decoding with reduced latency while relaxing the complexity constraint.}
Remember that in the decoding problem we know the optimal solution (i.e., \ac{ML} or \ac{MAP} decoders), however, due to the infeasible complexity for practical codes, we have to rely on sub-optimal decoders with a practical decoding cost (e.g., \ac{SPA} \ac{BP} decoder in the \ac{LDPC} decoding context).
To highlight the sub-optimality of the \ac{LDPC} \ac{BP} decoder in the short-length regime we refer to \cite[Fig.~4]{OSD1} and \cite[Fig.~10]{Buchberger_Journal}.
For short-length \ac{LDPC} codes, a huge performance gap (in $E_\mathrm{b}/N_0$) exists between \ac{BP} decoding and the \ac{ML} bound which can be estimated via an \ac{ML}-approaching \ac{OSD}.
Closing this performance gap is the main motive behind this work.\footnote{Note that all of the presented error-rate performance gains are attributed to the enhanced decoding algorithm itself (not to be confused with gains due to better code design).}

Ensemble decoding is a method to improve decoding performance by employing $L$ parallel independent \ac{BP} decoders each proposing a codeword estimate and then selecting the most likely candidate as the decoder output. \new{Two instances of ensemble decoding are augmented \ac{BP} \cite{FossorierAugmentedBP} and \ac{SBP} \cite{WehnSaturatedMinSum} decoding, where all possible combinations of saturated \ac{LLR} values in the $S$ least reliable positions of the received sequence are used as inputs to the constituent decoders. Another variant of ensemble decoding is \ac{MBBP} decoding \cite{Huber} (or \ac{BPL} decoding in the context of polar codes \cite{elkelesh2018belief}), where each \ac{BP} decoder uses a different decoding graph rather than a different input.} When the automorphism group of the code is known, identical constituent decoders decoding permuted versions of the channel output may be used, yielding so-called \ac{AED}. This has been successfully applied to high-density cyclic codes \cite{Hehn_MBBP_cyclic}, \ac{RM} codes \cite{rm_automorphism_ensemble_decoding} and polar codes \cite{PolarAutomorphisms_ISIT21}. Moreover, a sequential (rather than parallel) variant of automorphism-based decoding has been proposed in \cite{Dimnik_RRD_HDPC}.

For \ac{QC} \ac{LDPC} codes, however, the decoder equivariance phenomenon \cite{Chen_Cyclic_LDPC_AED} previously prevented successful application of \ac{AED}. We show that a small variation in the decoding Tanner graph is enough to exploit \ac{AED} with permutation vectors from the automorphism group of the considered \ac{QC} \ac{LDPC} code, i.e., quasi-cyclic shifts of the code symbols. Thus, our proposed decoding algorithm can be directly applied to standardized state-of-the-art \ac{QC} \ac{LDPC} codes without any special code design constraint (i.e., no changes on the encoder side, when compared to \cite{Chen_Cyclic_LDPC_AED, Zhang_Parallel_LDPC}). Standardized codes are usually flexible in codelength by specifying different protograph lifting factors and, thus, many receiver architectures already provide parallel hardware resources used only for large block-lengths. \ac{AED} may exploit these additional resources as independent parallel decoders and, thus, promises gains with minimal hardware overhead and low latency.

\section{Preliminaries}\label{sec:preliminaries}
\subsection{Structure of LDPC Codes}
LDPC codes were originally introduced by Gallager \cite{Gallager} as codes that could be conventionally represented by its corresponding $(M\times N)$ parity-check matrix $\mathbf{H} = \left [h_{ji}\right]_{M\times N}$, where $N$ is the number of \acp{VN} (i.e., also the code block-length) and $M$ represents the number of \acp{CN}. Therefore, the information bit block-length is $K = N-\text{rank}(\mathbf{H})$. Accordingly, the actual code rate\footnote{An actual code rate could be potentially higher than the so-called design rate $r_\mathrm{d} = (N-M)/N$.} is designated by $R_\mathrm{c} = K/N$. 
Additionally, there exists a corresponding graphical representation, namely the Tanner graph, where the \emph{bipartite} sets of nodes, namely, \acp{VN} and \acp{CN}, are connected according to $\mathbf{H}$ (i.e., a VN $v_i$ is connected to a \ac{CN} $c_j$ if $h_{ji} = 1$, with $i \in \left\{0,\dots, N-1\right\}$ and $j \in \left\{0,\dots, M-1\right\}$).

\subsection{BP Decoding}\label{ssec:bpdecoding}
Alongside \ac{LDPC} codes, Gallager introduced a suitable iterative decoding scheme \cite{Gallager} whose modified version is today known as the \ac{BP} algorithm (also known as \ac{SPA}). The algorithm passes messages, in form of extrinsic \acp{LLR}, along the edges of the Tanner graph. The result is an iterative update process at the \acp{VN} and \acp{CN}.
Each \ac{VN} can be interpreted as a repetition code and, thus, the update equation is
\begin{equation}\label{eq: vn_update}
    L_{v_i \rightarrow c_j} = L_{ch,i} + \sum_{j'\neq j}L_{c_{j'}\rightarrow v_i}
\end{equation}
where $L_{v_i \rightarrow c_j}$ is the outgoing message from the \ac{VN} $v_i$ to the \ac{CN} $c_j$,  $L_{ch,i}$ is the $i$-th channel output \ac{LLR} and $L_{c_{j}\rightarrow v_i}$ is the incoming message from the \ac{CN} $c_j$ to the \ac{VN} $v_i$. In contrast, each \ac{CN} can be seen as a single parity-check code which yields the update equation
\begin{equation}
    L_{c_j \rightarrow v_i} = 2\cdot \tanh^{-1}\left( \prod_{i'\neq i} \tanh \left(\frac{ L_{v_{i'}\rightarrow c_j}}{2} \right) \right).
\end{equation}
\new{In the final \ac{VN} calculation, all incoming messages are summed up to obtain the total \acp{LLR}. An implementation friendly variant of \ac{BP} is so-called layered decoding, where \acp{CN} are processed sequentially, incorporating the output of previous \acp{CN} already within the same iteration, resulting in faster convergence.}
For more details about \ac{LDPC} codes and \ac{BP} decoding, we refer the interested reader to \cite{densityEvol}.

\section{Code Symmetry vs. Decoder Symmetry}
The key aspect to enable \ac{AED} for \ac{LDPC} codes is the relationship between symmetries of the code and symmetries of the decoder. 

\subsection{Code Symmetry}
The permutation symmetries of a code $\mathcal{C}$ with length $N$ are given by its \textit{automorphism group} $\operatorname{Aut}(\mathcal{C})$. It is defined as the set of codeword symbol permutations that map every codeword onto another (not necessarily different) codeword:
\begin{equation}
    \operatorname{Aut}(\mathcal{C}) = \left\{\pi \in \mathcal{S}_N: \pi(\mathbf{c}) \in \mathcal{C} \; \forall \mathbf{c} \in \mathcal{C}\right\},
\end{equation}
where $\mathcal{S}_N$ denotes the symmetric group of $N$ elements \cite{macwilliams77}.

\begin{figure} 
	\centering
	\resizebox{\columnwidth}{!}{\input{tikz/AED.tikz}}
	\caption{\footnotesize Block diagram of automorphism ensemble decoding (AED) of a noisy codeword $\mathbf{y}$ with $L$ identical BP-based constituent decoders.}
	\label{fig:Block_Diag}
	\vspace{-0.25cm}
\end{figure}
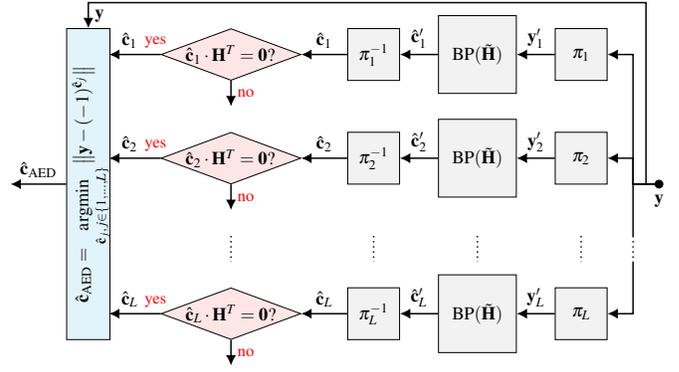

\subsection{Automorphism Ensemble Decoding}
Let $\operatorname{Dec}(\cdot): \mathcal{Y}^N \to \mathcal{C}$ denote the decoding function, where $\mathcal{Y}$ is the set of possible channel outputs. For instance $\mathcal{Y}$ is the set of real numbers $\mathbb{R}$ in case of the \ac{BI-AWGN} channel. \ac{AED} \cite{rm_automorphism_ensemble_decoding} attempts to decode multiple, differently permuted versions of the noisy codeword $\mathbf{y}$, using a subset $\mathcal{P} \subseteq \operatorname{Aut}(\mathcal{C})$ of $L$ permutations. Each permutation $\pi_i\in\mathcal{P}$ contributes one codeword candidate
\begin{equation}
    \hat{\mathbf{c}}_j=\pi_j^{-1}(\operatorname{Dec}(\pi_j(\mathbf{y}))),
\end{equation}
from which the final \ac{AED} codeword estimate is chosen using the \ac{ML} criterion
\begin{equation}
    \hat{\mathbf{c}}_\mathrm{AED} = \argmax_{\hat{\mathbf{c}}_j, j\in\{1,2,\dots,L\}} P(\hat{\mathbf{c}}_j|\mathbf{y}).
\end{equation}
Fig. \ref{fig:Block_Diag} shows the block diagram of \ac{AED} with constituent \ac{BP} decoders and a selection criterion based on Euclidean distance, which is the \ac{ML} criterion for the \ac{BI-AWGN} channel. It is easy to see that permuted decoding with permutation $\pi$ and parity-check matrix $\mathbf{H}$ is identical to decoding on the column-permuted parity-check matrix
\begin{equation}
    \mathbf{H}' = \pi^{-1}(\mathbf{H}).
\end{equation}
Therefore, \ac{AED} with \ac{BP} decoders is a special case of \ac{MBBP}~\cite{Huber}, where the used $\mathbf{H}$-matrices only differ by column permutations out of the automorphism group of the code.
In this work, we use the notation AED-$L$ to denote an \ac{AED} with ensemble size $L$.

\subsection{Decoder Symmetry}
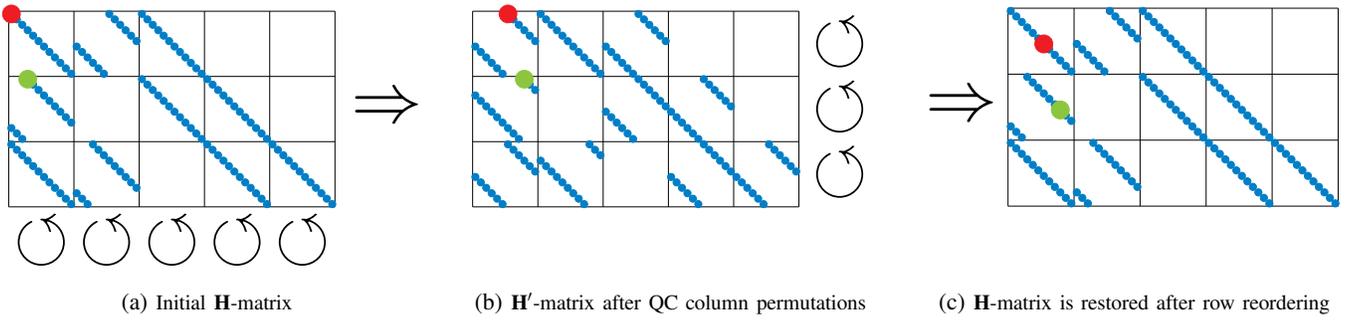
\begin{figure*}[t]
   \begin{subfigure}[b]{0.32\linewidth}
	\resizebox{1\columnwidth}{!}{
	\input{tikz/Original_H.tikz}
	} 
	\caption{\footnotesize Initial $\mathbf{H}$-matrix} 
	\label{fig:initH}
            \end{subfigure}
       \hfill     \begin{subfigure}[b]{0.32\linewidth}
	\resizebox{1\columnwidth}{!}{
	\input{tikz/cols_permuted_H.tikz}
	} 
	\caption{\footnotesize \new{$\mathbf{H}'$}-matrix after QC column permutations}
	\label{fig:H_dash}
    \end{subfigure}       
    \hfill
    \begin{subfigure}[b]{0.32\linewidth}
		\resizebox{1\columnwidth}{!}{
		\input{tikz/rows_permuted_H.tikz}
		} 
		\caption{\footnotesize $\mathbf{H}$-matrix is restored after row reordering}
		\label{fig:H_double_dash}
    \end{subfigure}	          
  
\caption{\footnotesize Decoder equivariance: The conventional parity-check matrix of a \ac{QC} \ac{LDPC} code absorbs quasi-cyclic permutations of the columns. While the highlighted elements change positions, the overall $\mathbf{H}$-matrix remains unchanged.} \label{fig:AEDLDPC}
 \vspace{-0.25cm}
 \end{figure*}

Not all permutations are useful for \ac{AED}, as they result in the same codeword candidates. To analyze this, we say a decoder is \textit{equivariant} to a permutation $\pi$, if permuting its input $\mathbf{y}$ is the same as permuting its output $\hat{\mathbf{c}}$. In other words, the permutation operation commutes with the decoding operation:
\begin{equation}
    \operatorname{Dec}\left(\pi(\mathbf{y})\right) = \pi\left(\operatorname{Dec}(\mathbf{y})\right) \quad \forall \mathbf{y}\in\mathcal{Y}^N.
\end{equation}
We say that $\pi$ is \textit{absorbed} by $\operatorname{Dec}(\cdot)$ \cite{rm_automorphism_ensemble_decoding}. Each absorbed permutation $\pi$ induces sets $\left\{\pi\circ\sigma|\sigma\in\operatorname{Aut}(\mathcal{C})\right\}$ of equivalent automorphisms. Let $\pi,\sigma_1 \in \operatorname{Aut}(\mathcal{C})$ and $\pi$ be absorbed by $\operatorname{Dec}(\cdot)$. Then the codeword estimate from $\sigma_2=\pi\circ\sigma_1$ is 
\begin{align}
\sigma_2^{-1}\left(\operatorname{Dec}\left(\sigma_2(\mathbf{y})\right)\right) &=
\sigma_1^{-1}\left(\pi^{-1}\left(\operatorname{Dec}\left(\pi\left(\sigma_1(\mathbf{y})\right)\right)\right)\right)\nonumber\\
&= \sigma_1^{-1}\left(\operatorname{Dec}(\sigma_1\left(\mathbf{y})\right)\right),
\end{align}
i.e., equivalent permutations $\sigma_1 \sim \sigma_2$ always result in the same codeword candidate under permuted decoding. It can be shown that equivalent permutations form equivalence classes which are themselves subgroups of $\operatorname{Aut}(\mathcal{C})$ \cite{pillet2021automorphismclassification}.  Therefore, decoder symmetries reduce the number of equivalence classes and, thus, also reduce the number of usable automorphisms for an ensemble decoder.

\subsection{Quasi-Cyclic Codes and Decoders}
A code $\mathcal{C}$ of length $N=nZ$ is called quasi-cyclic, if all permutations of the form
\begin{equation}
    \pi_{d,Z}(i) = \begin{cases}
i+d-Z & \text{if }i \text{ mod } Z + d \ge Z\\
i+d & \text{else}\\
\end{cases}
\end{equation}
with $0\le d < Z$ are automorphisms of $\mathcal{C}$. Therefore, $\operatorname{Aut}(\mathcal{C})$ is at least the quasi-cyclic group of size $Z$
\begin{equation}
    \mathcal{Q}_Z = \left\{\pi_{d,Z}:\quad d=0,1,\dots, Z-1\right\}.
\end{equation}
Prominent representatives of the class of \ac{QC} codes are \ac{QC} \ac{LDPC} codes \cite{QCfossorier}. However, in this case, the \ac{QC} property mainly serves the ease of construction and implementation. A \ac{QC} \ac{LDPC} code is characterised by its parity-check matrix being expanded from a so-called protograph by a lifting factor $Z$. \new{In the lifting process, the elements of the protograph matrix are replaced by circulant submatrices of size $Z\times Z$. 
Their encoding can be thus realised by a set of shift registers, with the linear complexity with respect to the total code length \cite{QCefficientEncoding}}. Moreover, various code lengths can be easily realized from a single protograph using different lifting factors $Z$.

The $\left(Zm\times Zn\right)$ QC LDPC code $\mathbf{H}$-matrix~can~be~written~as  $$\mathbf{H}=\left[\begin{array}{cccc}
	\mathbf{H}_{0,0} & \mathbf{H}_{0,1} & \cdots & \mathbf{H}_{0,n-1}\\
	\mathbf{H}_{1,0} & \mathbf{H}_{1,1} & \cdots & \mathbf{H}_{1,n-1}\\
	\vdots & \vdots & \ddots & \vdots \\
	\mathbf{H}_{m-1,0} & \mathbf{H}_{m-1,1} & \cdots & \mathbf{H}_{m-1,n-1}
\end{array}\right],$$
where submatrices $\mathbf{H}_{i,j}$ of size $(Z\times Z)$ are circulant.

It can be seen that both the rows and columns of the parity-check matrix fulfill the quasi-cyclic property. While quasi-cyclicity of the columns creates the automorphism group, quasi-cyclic rows result in decoder equivariance to these permutations. As shown in \cite{Chen_Cyclic_LDPC_AED}, permuted \ac{BP} decoding (with the permutation $\pi_{d,Z}$) is equivalent to \ac{BP} decoding on the column-permuted parity-check matrix
\begin{equation}
    \new{\mathbf{H}'} = \pi_{d,Z}^{-1}(\mathbf{H}) = \pi_{Z-d,Z}(\mathbf{H}), 
\end{equation}
which is just a row-permuted version of $\mathbf{H}$ (as visualized in Fig.~\ref{fig:AEDLDPC}). For that reason, the column-permuted parity-check matrix shows exactly the same decoding behaviour in a flooding decoder as the original parity-check matrix. \new{The same applies to layered decoding with a regular schedule, as the permutation only affects sets of independent checks.}
Therefore, \ac{AED} using the standard $\mathbf{H}$-matrices and \ac{QC} permutations does not result in any performance gain for \ac{QC} \ac{LDPC} codes.

\section{Breaking Decoder Symmetry}
To successfully apply \ac{AED} to \ac{QC} \ac{LDPC} codes, one can either design codes whose automorphism group is larger than $\mathcal{Q}_Z$ (such as the codes proposed in \cite{Chen_Cyclic_LDPC_AED}), or break the symmetry group of the constituent decoders to be smaller than $\mathcal{Q}_Z$. We propose the latter method, as it does not require a specific code design and hence is compatible with standardized \ac{QC} \ac{LDPC} codes. We still apply the conventional \ac{BP} decoding algorithm as introduced in Sec. \ref{ssec:bpdecoding}, however, on a different Tanner graph $(\tilde{\mathbf{H}})$ which is not quasi-cyclic. As the original Tanner graph is designed to optimize the performance under \ac{BP} decoding, it serves as a natural starting point. We propose three methods to break the symmetry by modifying the original Tanner graph (i.e., three methods of finding the $\tilde{\mathbf{H}}$-matrix):

\begin{enumerate}
    \item \textit{Row operations}: 
    Elementary row operations on the parity-check matrix do not change the code but result in different Tanner graphs. In the case of binary codes, the only interesting row operation is adding a row onto another.
    \item \textit{Adding Auxiliary Checks (``overcomplete'')}: 
    One can add a single or multiple auxiliary checks to the parity-check matrix. The added checks should be linear combinations of the original checks, such that the resulting matrix is still a valid, overcomplete, parity-check matrix $\mathbf{\tilde{H}}$.
    \item \textit{Removing Checks (``undercomplete'')}: 
     We propose to remove some checks, resulting in an undercomplete parity-check matrix $\mathbf{\tilde{H}}$, which strictly-speaking means changing the considered code. This matrix belongs to a code that contains, besides the codewords of the original code, further invalid codewords. \ac{AED} must detect when a constituent decoder converged to such an invalid codeword. Therefore, the original $\mathbf{H}$-matrix is used to check ``code membership'' and only \emph{valid} candidates are included in the ML-in-the-list selection, as shown in Fig.~\ref{fig:Block_Diag}.
\end{enumerate}

Note that all proposed methods operate on the full, lifted parity-check matrix rather than the protograph. 

\section{Results}
\subsection{Error Rate Performance}
\begin{figure} [t]
	\centering
	\resizebox{\columnwidth}{!}{\input{tikz/methods.tikz}}
	\caption{\footnotesize Comparison of the proposed parity-check matrix modifications for the ($N=132$, $K=66$) 5G LDPC code. All iterative decoders use 32 iterations.}
	\label{fig:methods}
\end{figure}
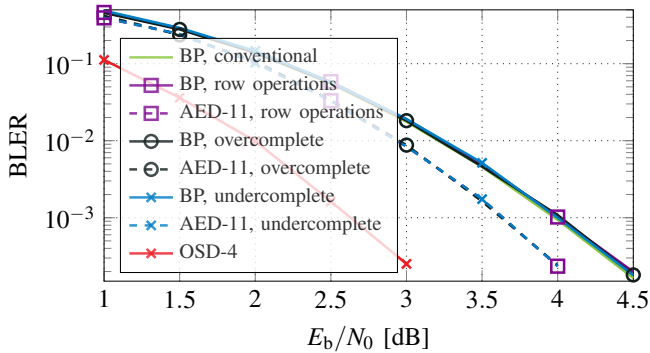

We evaluate the performance of the proposed methods on various \ac{QC} \ac{LDPC} codes from communications standards. Table~\ref{tab:codes} lists the used code parameters. All \ac{BP} decoders are implemented as floating point \ac{SPA} with flooding schedule and are simulated using an \ac{BI-AWGN} channel with \ac{BPSK} modulation. If available, we also plot the \ac{ML} performance of the corresponding code \cite{kldatabase} or, if computationally feasible, an approximation using \ac{OSD} \new{\cite{OSD}}, where OSD-$t$ denotes \ac{OSD} with order $t$.

\begin{table}[h]
\caption{\footnotesize{Parameters of the considered LDPC codes}}
\centering\begin{tabular}{c|cccc}
    Code & $N$ & $K$ & $R_c$ & $Z$ \\
    \hline
    802.11n & 648 & 540 & $\nicefrac{5}{6}$ & 27 \\
    5G, BG 2 & 132 & 66 & $\nicefrac{1}{2}$ & 11 \\
    5G, BG 2 & 264 & 132 & $\nicefrac{1}{2}$ & 22 \\
    CCSDS & 128 & 64 & $\nicefrac{1}{2}$ & 16 \\
    CCSDS & 256 & 128 & $\nicefrac{1}{2}$ & 32 \\
\end{tabular}\label{tab:codes}
\end{table}

We first compare the three proposed methods in their error-rate performance using the (132,66) 5G LDPC code. While there exist infinite ways to combine and extend the alteration methods, we only change, add or remove a single check to demonstrate the capability of the method. 
The first modification adds check 0 onto check 1, i.e., changing the check 1. For the overcomplete case, we appended an additional check which is the mod-2-sum of checks 51, 53, 58 and 71 (counting from~0). \new{This combination was chosen randomly, however, with the constraint that the number of involved variable nodes is relatively low. In this case, the degree of the auxiliary check is 11.} Lastly, we use an undercomplete $\tilde{\mathbf{H}}$-matrix where the zeroth check has been deleted. 
\new{Fig. \ref{fig:methods} shows the \ac{BLER} performance of the proposed methods.} To fully exploit the capability of \ac{AED}, we use all $Z=11$ available quasi-cyclic permutations, i.e., $L=Z$.

It can be seen that while all modifications slightly degrade the performance compared to the original parity-check matrix, in all cases, the ensemble of $L=Z$ decoders outperforms this baseline decoder by a significant margin. To our surprise, all methods show virtually identical gains. Therefore, in the following, we focus on the undercomplete $\tilde{\mathbf{H}}$ variant, as its implementation is the easiest. In fact, a conventional decoder may be used with a single check being deactivated. 

In Fig. \ref{fig:wifi_results} we show results for the (648,540) Wi-Fi code. The \ac{AED} uses an undercomplete $\tilde{\mathbf{H}}$-matrix with the zeroth check removed. Even though the code is of moderate length and, thus, the gap to its \ac{ML} performance is already less than 1 dB at a \ac{BLER} of $10^{-4}$, the proposed ensemble decoder (AED) produces gains of approximately 0.2 dB. \new{We also show results for a check-node layered decoding with 16 iterations, where even larger gains are achieved by \ac{AED}.}

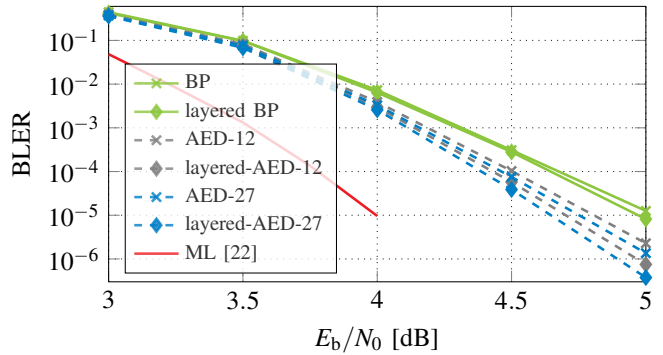
\begin{figure} [t]
	\centering
	\resizebox{\columnwidth}{!}{\input{tikz/wifi_results.tikz}}
	\caption{\footnotesize Results of the Wi-Fi 802.11n code. The flooding BP decoders use 32 iterations, while the check-node layered decoders use 16 iterations. An undercomplete parity-check matrix $\tilde{\mathbf{H}}$ is used in the AED simulations.}
	\label{fig:wifi_results}
\end{figure}

In Fig. \ref{fig:5g132_results} we show results for different rate-half 5G LDPC codes. Again, the \ac{AED} uses an ensemble of $L=Z$ \ac{BP} decoders using an undercomplete parity-check matrix ($\tilde{\mathbf{H}}$-matrix) with the zeroth check removed. For both block lengths, at a \ac{BLER} of $10^{-3}$, we see gains of 0.3 dB and 0.2 dB, respectively.

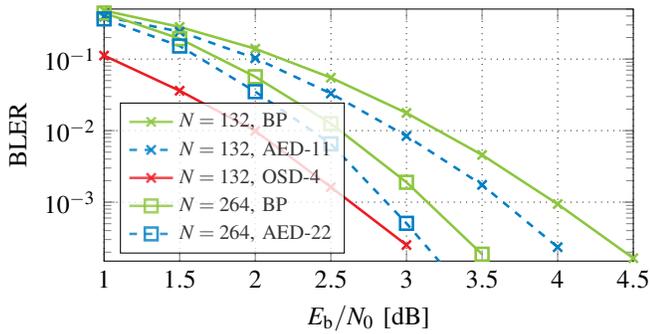
\begin{figure} 
	\centering
	\resizebox{\columnwidth}{!}{\input{tikz/5g_132_results.tikz}}
	\vspace{-0.3cm}
	\caption{\footnotesize Results for rate-half ($N$, $K=N/2$) 5G LDPC codes. All BP decoders use 32 iterations each, AED uses an  undercomplete $\tilde{\mathbf{H}}$ matrix.}
	\label{fig:5g132_results}
\end{figure}

In Fig. \ref{fig:ccsds_results}, we plot results for rate-half CCSDS codes and the same \ac{AED} parameters. For both block lengths, \ac{AED} achieves a gain of approximately 0.2 dB at a \ac{BLER} of $10^{-4}$ when compared to conventional \ac{BP} decoding (i.e., gain due to the enhanced decoding algorithm). \new{Moreover, we compare to \ac{SBP} with the same number of constituent decoders, i.e, $S=4$ for $Z=16$ and $S=5$ for $Z=32$. We see that the performance of \ac{AED} is very similar to that of \ac{SBP}, while in the higher SNR regime, \ac{AED} can slightly outperform \ac{SBP}. Note that to make the comparison fair, we use full \ac{SPA} decoders rather than the min-sum approximation as proposed in \cite{WehnSaturatedMinSum}.}

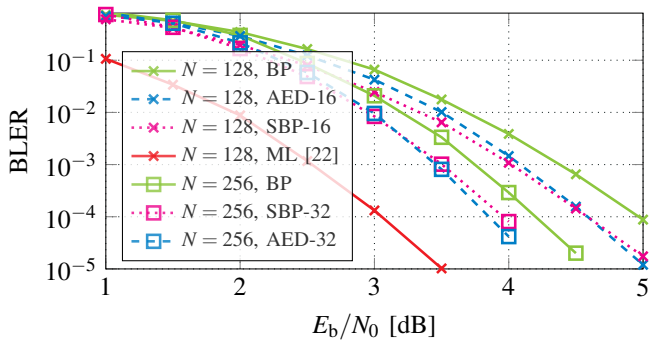
\begin{figure} 
	\centering
	\resizebox{\columnwidth}{!}{\input{tikz/ccsds_results.tikz}}
	\vspace{-0.15cm}
	\caption{\footnotesize Results for rate-half ($N$, $K=N/2$) CCSDS codes. All BP decoders use 32 iterations each, AED uses an  undercomplete $\tilde{\mathbf{H}}$ matrix.}
	\label{fig:ccsds_results}
	\vspace{-0.4cm}
\end{figure}

\new{
\subsection{Latency}
Compared to conventional \ac{BP} decoding, \ac{AED} has a lower worst case latency, as the number of iterations required to reach the same \ac{BLER} performance is reduced. For example, for the (128,64) CCSDS code, 256 single BP iterations are required to match the AED performance with 32 iterations. Compared to \ac{SBP}, a lower worst case latency is expected, as the dynamic preprocessing of finding the least reliable positions is not required. In terms of average latency, i.e., with early stopping, we find that \ac{AED} with an average of 3.4 iterations significantly improves over \ac{SBP} requiring 28.5 iterations at $E_\mathrm{b}/N_0 = 4~\text{dB}$. This is because in \ac {SBP}, only an average of 1.49 of the 16 constituent decoders converge at all. Note that for this analysis, in case of AED we require all constituent decoders to be done before an overall decoding result is available, while for \ac{SBP} the decoding is stopped once 3 decoders have converged, as proposed in \cite{WehnSaturatedMinSum}.
}

\section{Conclusion}\label{sec:conc}
In this work we demonstrated that breaking the symmetry in the parity-check matrix on the decoder side can enable \ac{AED} for \ac{QC} \ac{LDPC} codes. Even without any optimization of how exactly the parity-check matrix is altered, consistent gains between 0.2 dB and 0.3 dB over conventional \ac{BP} decoding could be achieved. Larger gains are indeed expected when further optimizations are applied. Additionally, many more ways of breaking the decoder symmetry remain to be explored, \new{such as non-standard schedules, which has been already successfully applied to polar codes in \cite{CRC_BPL_ISIT20}}.
\bibliographystyle{IEEEtran}
\bibliography{references}
\end{NoHyper}
\end{document}

%% file: corporateColours.tex
\definecolor{mittelblau}{RGB}{0, 126, 198}
\definecolor{violettblau}{cmyk}{0.9, 0.6, 0, 0}
\definecolor{rot}{RGB}{238, 28 35}
\definecolor{apfelgruen}{RGB}{140, 198, 62}
\definecolor{gelb}{RGB}{255, 229, 0}
\definecolor{orange}{RGB}{244, 111, 33}
\definecolor{pink}{RGB}{237, 0, 140}
\definecolor{lila}{RGB}{128, 10, 145}
\definecolor{hellgrau}{RGB}{224, 224, 224}
\definecolor{mittelgrau}{RGB}{128, 128, 128}
\definecolor{dunkelgrau}{RGB}{80,80,80}
\definecolor{anthrazit}{RGB}{19, 31, 31}
\definecolor{darkgreen}{RGB}{34,139,34}
\definecolor{aqua}{RGB}{0, 255, 255}

%% file: nodesAndStyles.tex
\tikzset{
       vnd/.style={
        shape=circle,
        fill=black,
        draw,
        inner sep=0pt,
        minimum size=0.2cm},
        cnd/.style={
        shape=rectangle,
        fill=white,
        draw,
        minimum width=0.05mm,
        minimum height = 0.05mm}, 
         vndR/.style={
        shape=circle,
        fill=red,
        draw,
        inner sep=0pt,
        minimum size=0.2cm},
        cndR/.style={
        shape=rectangle,
        fill=white,
        draw=red,
        minimum width=0.05mm,
        minimum height = 0.05mm}
}

%% file: macros.tex
\newcommand{\argmax}{\mathop{\mathrm{argmax}}}



%% file: tikz/AED.tikz
\begin{tikzpicture}
\tikzset{
edge/.style = {thick,black},
mydiamond/.style={draw, diamond, aspect=2.7,text width=1.75cm, inner sep=0pt,  fill=white!90!red},
BPrectangle/.style={rectangle, draw, minimum size=1.5cm, fill=white!90!gray},
intrect/.style={rectangle, draw, minimum size=1cm, fill=white!90!gray}
}

\tikzstyle{conn} = [-{Latex[length=2mm,width=2mm]}];

\node[draw,shape=circle, fill=black, inner sep=0pt,minimum size=0.15cm, label=below:{$\mathbf{y}$}] (Lch) at (2, 0.5) {};

\draw [edge, dotted] (-1.5,-0.5)--(-1.5,-1);
\draw [edge] (Lch)--(1.5,0.5)--(1.5,-0.5);
\draw [edge, dotted] (1.5,-0.5)--(1.5,-1);
\draw [edge, black,-, dotted] (-6.25,-0.5)--(-6.25,-1);
\draw [edge, black,-, dotted] (-3.5,-0.5)--(-3.5,-1);
\draw [edge, black,-, dotted] (.5,-0.5)--(.5,-1);

\node[BPrectangle] (BP1) at (-1.5, 3) {$\operatorname{BP}(\tilde{\mathbf{H}})$};
\node[intrect] (int1) at (.5, 3) {$ \pi_1 $};
\node[intrect] (deint1) at (-3.5, 3) {$ \pi_1^{-1} $};
\draw [edge,conn] (Lch)--(1.5,0.5)--(1.5,3)--(int1);
\node[mydiamond] (cond1) at (-6.25, 3) {$\hat{\mathbf{c}}_1\cdot \mathbf{H}^T=\mathbf{0}?$};
\draw [edge,conn](cond1) to node[right,red] {\small{no}} (-6.25,2);
\path [edge](cond1) to node[above,red,xshift=.1cm] {\small{yes}} (-8,3);
\draw [edge,conn](cond1) to node[above,xshift=-.1cm] {$\hat{\mathbf{c}}_1$} (-8.6,3);

\node[BPrectangle] (BP2) at (-1.5, 1) {$\operatorname{BP}(\tilde{\mathbf{H}})$};
\node[intrect] (int2) at (.5, 1) {$ \pi_2 $};
\node[intrect] (deint2) at (-3.5, 1) {$ \pi_2^{-1} $};
\draw [edge,conn] (Lch)--(1.5,0.5)--(1.5,1)--(int2);
\node[mydiamond] (cond2) at (-6.25, 1) {$\hat{\mathbf{c}}_2\cdot \mathbf{H}^T=\mathbf{0}?$};
\draw [edge,conn](cond2) to node[right,red] {\small{no}} (-6.25,0);
\path [edge](cond2) to node[above,red,xshift=.1cm] {\small{yes}} (-8,1);
\draw [edge,conn](cond2) to node[above,xshift=-.1cm] {$\hat{\mathbf{c}}_2$} (-8.6,1);

\node[BPrectangle] (BP3) at (-1.5, -2) {$\operatorname{BP}(\tilde{\mathbf{H}})$};
\node[intrect] (int3) at (.5, -2) {$ \pi_L $};
\node[intrect] (deint3) at (-3.5, -2) {$ \pi_L^{-1} $};
\draw [edge,conn] (1.5,-1)--(1.5,-2)--(int3);
\node[mydiamond] (cond3) at (-6.25, -2) {$\hat{\mathbf{c}}_L\cdot \mathbf{H}^T=\mathbf{0}?$};
\draw [edge,conn](cond3) to node[right,red] {\small{no}} (-6.25,-3);
\path [edge](cond3) to node[above,red,xshift=.1cm] {\small{yes}} (-8,-2);
\draw [edge,conn](cond3) to node[above,xshift=-.1cm] {$\hat{\mathbf{c}}_L$} (-8.6,-2);

\draw [edge,conn](int1) to node[above] {$\mathbf{y}'_1$} (BP1);
\draw [edge,conn](int2) to node[above] {$\mathbf{y}'_2$} (BP2);
\draw [edge,conn](int3) to node[above] {$\mathbf{y}'_L$} (BP3);

\draw [edge,conn](BP1) to node[above] {$\hat{\mathbf{c}}'_1$} (deint1);
\draw [edge,conn](BP2) to node[above] {$\hat{\mathbf{c}}'_2$} (deint2);
\draw [edge,conn](BP3) to node[above] {$\hat{\mathbf{c}}'_L$} (deint3);

\draw [edge,conn](deint1) to node[above] {$\hat{\mathbf{c}}_1$} (cond1);
\draw [edge,conn](deint2) to node[above] {$\hat{\mathbf{c}}_2$} (cond2);
\draw [edge,conn](deint3) to node[above] {$\hat{\mathbf{c}}_L$} (cond3);


\node[rectangle, draw, minimum width=6cm, minimum height=0.3cm, text height=0.3cm, text depth=0.3cm,text centered,rotate=90, fill=white!90!cyan] (decide) at (-9, 0.5) {$\hat{\mathbf{c}}_\mathrm{AED}=\underset{\hat{\mathbf{c}}_{j},j\in\left\{ 1,\dots,L\right\} }{\mathrm{argmin}}\left\Vert \mathbf{y}-(-1)^{\hat{\mathbf{c}}_{j}}\right\Vert $};

\draw [edge,conn] (Lch)--(1.75,0.5)--(1.75,4)--(-9,4)to node[right] {$\mathbf{y}$}(decide);
\draw [edge,conn](decide) to node[above] {$\hat{\mathbf{c}}_\mathrm{AED}$} (-10.5,0.5);

\end{tikzpicture}

%% file: tikz/Original_H.tikz
\begin{tikzpicture}

\begin{axis}[%
axis equal,
scale only axis,
axis x line=none,
axis y line=none,
xmin=-2,
xmax=76,
y dir=reverse,
ymin=-2,
ymax=46,
]
\addplot [only marks, color=mittelblau, mark=*, mark size=2pt]
  table[row sep=crcr]{%
1	1\\
1	22\\
1	25\\
2	2\\
2	23\\
2	26\\
3	3\\
3	24\\
3	27\\
4	4\\
4	13\\
4	28\\
5	5\\
5	14\\
5	29\\
6	6\\
6	15\\
6	30\\
7	7\\
7	16\\
7	31\\
8	8\\
8	17\\
8	32\\
9	9\\
9	18\\
9	33\\
10	10\\
10	19\\
10	34\\
11	11\\
11	20\\
11	35\\
12	12\\
12	21\\
12	36\\
13	7\\
13	34\\
14	8\\
14	35\\
15	9\\
15	36\\
16	10\\
16	25\\
17	11\\
17	26\\
18	12\\
18	27\\
19	1\\
19	28\\
20	2\\
20	29\\
21	3\\
21	30\\
22	4\\
22	31\\
23	5\\
23	32\\
24	6\\
24	33\\
25	1\\
25	13\\
26	2\\
26	14\\
27	3\\
27	15\\
28	4\\
28	16\\
29	5\\
29	17\\
30	6\\
30	18\\
31	7\\
31	19\\
32	8\\
32	20\\
33	9\\
33	21\\
34	10\\
34	22\\
35	11\\
35	23\\
36	12\\
36	24\\
37	13\\
37	25\\
38	14\\
38	26\\
39	15\\
39	27\\
40	16\\
40	28\\
41	17\\
41	29\\
42	18\\
42	30\\
43	19\\
43	31\\
44	20\\
44	32\\
45	21\\
45	33\\
46	22\\
46	34\\
47	23\\
47	35\\
48	24\\
48	36\\
49	25\\
50	26\\
51	27\\
52	28\\
53	29\\
54	30\\
55	31\\
56	32\\
57	33\\
58	34\\
59	35\\
60	36\\
};

\addplot [only marks, color=rot, mark=*, mark size=5.0pt]
  table[row sep=crcr]{%
1	1\\
};

\addplot [only marks, color=apfelgruen, mark=*, mark size=5.0pt]
  table[row sep=crcr]{%
4	13\\
};

\foreach \x in {0,...,5}
{
\edef\temp{\noexpand\draw [black] (axis cs:\x*12+.5,0.5) -- (axis cs:\x*12+.5,36.5);}
    \temp
}
\foreach \y in {0,...,3}
{
\edef\temp{\noexpand\draw [black] (axis cs:0.5,\y*12+0.5) -- (axis cs:60.5,\y*12+0.5);}
    \temp
}

\node[] at (axis cs:6.5,43) {\fontsize{40}{60}\color{black}$\circlearrowleft$};
\node[] at (axis cs:18.5,43) {\fontsize{40}{60}\color{black}$\circlearrowleft$};
\node[] at (axis cs:30.5,43) {\fontsize{40}{60}\color{black}$\circlearrowleft$};
\node[] at (axis cs:42.5,43) {\fontsize{40}{60}\color{black}$\circlearrowleft$};
\node[] at (axis cs:54.5,43) {\fontsize{40}{60}\color{black}$\circlearrowleft$};

\node[] at (axis cs:70,18.5) {\fontsize{40}{60}\color{black}$\Rightarrow$};
\draw[draw=white] (axis cs:-2,-2) rectangle (axis cs:76,49);

\end{axis}



\end{tikzpicture}%

%% file: tikz/cols_permuted_H.tikz
\begin{tikzpicture}

\begin{axis}[%
scale only axis,
axis equal,
axis x line=none,
axis y line=none,
xmin=-2,
xmax=76,
y dir=reverse,
ymin=-2,
ymax=46,
]
\addplot [only marks, color=mittelblau, mark=*, mark size=2pt]
  table[row sep=crcr]{%
1	7\\
1	16\\
1	31\\
2	8\\
2	17\\
2	32\\
3	9\\
3	18\\
3	33\\
4	10\\
4	19\\
4	34\\
5	11\\
5	20\\
5	35\\
6	12\\
6	21\\
6	36\\
7	1\\
7	22\\
7	25\\
8	2\\
8	23\\
8	26\\
9	3\\
9	24\\
9	27\\
10	4\\
10	13\\
10	28\\
11	5\\
11	14\\
11	29\\
12	6\\
12	15\\
12	30\\
13	1\\
13	28\\
14	2\\
14	29\\
15	3\\
15	30\\
16	4\\
16	31\\
17	5\\
17	32\\
18	6\\
18	33\\
19	7\\
19	34\\
20	8\\
20	35\\
21	9\\
21	36\\
22	10\\
22	25\\
23	11\\
23	26\\
24	12\\
24	27\\
25	7\\
25	19\\
26	8\\
26	20\\
27	9\\
27	21\\
28	10\\
28	22\\
29	11\\
29	23\\
30	12\\
30	24\\
31	1\\
31	13\\
32	2\\
32	14\\
33	3\\
33	15\\
34	4\\
34	16\\
35	5\\
35	17\\
36	6\\
36	18\\
37	19\\
37	31\\
38	20\\
38	32\\
39	21\\
39	33\\
40	22\\
40	34\\
41	23\\
41	35\\
42	24\\
42	36\\
43	13\\
43	25\\
44	14\\
44	26\\
45	15\\
45	27\\
46	16\\
46	28\\
47	17\\
47	29\\
48	18\\
48	30\\
49	31\\
50	32\\
51	33\\
52	34\\
53	35\\
54	36\\
55	25\\
56	26\\
57	27\\
58	28\\
59	29\\
60	30\\
};

\addplot [only marks, color=rot, mark=*, mark size=5.0pt]
  table[row sep=crcr]{%
7	1\\
};

\addplot [only marks, color=apfelgruen, mark=*, mark size=5.0pt]
  table[row sep=crcr]{%
10	13\\
};

\foreach \x in {0,...,5}
{
\edef\temp{\noexpand\draw [black] (axis cs:\x*12+.5,0.5) -- (axis cs:\x*12+.5,36.5);}
    \temp
}
\foreach \y in {0,...,3}
{
\edef\temp{\noexpand\draw [black] (axis cs:0.5,\y*12+0.5) -- (axis cs:60.5,\y*12+0.5);}
    \temp
}

\node[] at (axis cs:68,6.5) {\fontsize{40}{60}\color{black}$\circlearrowleft$};
\node[] at (axis cs:68,18.5) {\fontsize{40}{60}\color{black}$\circlearrowleft$};
\node[] at (axis cs:68,30.5) {\fontsize{40}{60}\color{black}$\circlearrowleft$};

\draw[draw=white] (axis cs:-2,-2) rectangle (axis cs:76,49);
\end{axis}



\end{tikzpicture}%

%% file: tikz/rows_permuted_H.tikz
\begin{tikzpicture}

\begin{axis}[%
axis equal,
scale only axis,
axis x line=none,
axis y line=none,
xmin=-15,
xmax=63,
y dir=reverse,
ymin=0,
ymax=46,
]
\addplot [only marks, color=mittelblau, mark=*, mark size=2pt]
  table[row sep=crcr]{%
1	1\\
1	22\\
1	25\\
2	2\\
2	23\\
2	26\\
3	3\\
3	24\\
3	27\\
4	4\\
4	13\\
4	28\\
5	5\\
5	14\\
5	29\\
6	6\\
6	15\\
6	30\\
7	7\\
7	16\\
7	31\\
8	8\\
8	17\\
8	32\\
9	9\\
9	18\\
9	33\\
10	10\\
10	19\\
10	34\\
11	11\\
11	20\\
11	35\\
12	12\\
12	21\\
12	36\\
13	7\\
13	34\\
14	8\\
14	35\\
15	9\\
15	36\\
16	10\\
16	25\\
17	11\\
17	26\\
18	12\\
18	27\\
19	1\\
19	28\\
20	2\\
20	29\\
21	3\\
21	30\\
22	4\\
22	31\\
23	5\\
23	32\\
24	6\\
24	33\\
25	1\\
25	13\\
26	2\\
26	14\\
27	3\\
27	15\\
28	4\\
28	16\\
29	5\\
29	17\\
30	6\\
30	18\\
31	7\\
31	19\\
32	8\\
32	20\\
33	9\\
33	21\\
34	10\\
34	22\\
35	11\\
35	23\\
36	12\\
36	24\\
37	13\\
37	25\\
38	14\\
38	26\\
39	15\\
39	27\\
40	16\\
40	28\\
41	17\\
41	29\\
42	18\\
42	30\\
43	19\\
43	31\\
44	20\\
44	32\\
45	21\\
45	33\\
46	22\\
46	34\\
47	23\\
47	35\\
48	24\\
48	36\\
49	25\\
50	26\\
51	27\\
52	28\\
53	29\\
54	30\\
55	31\\
56	32\\
57	33\\
58	34\\
59	35\\
60	36\\
};

\addplot [only marks, color=rot, mark=*, mark size=5.0pt]
  table[row sep=crcr]{%
7	7\\
};

\addplot [only marks, color=apfelgruen, mark=*, mark size=5.0pt]
  table[row sep=crcr]{%
10	19\\
};

\foreach \x in {0,...,5}
{
\edef\temp{\noexpand\draw [black] (axis cs:\x*12+.5,0.5) -- (axis cs:\x*12+.5,36.5);}
    \temp
}
\foreach \y in {0,...,3}
{
\edef\temp{\noexpand\draw [black] (axis cs:0.5,\y*12+0.5) -- (axis cs:60.5,\y*12+0.5);}
    \temp
}
\node[] at (axis cs:-8,18.5) {\fontsize{40}{60}\color{black}$\Rightarrow$};

\draw[draw=white] (axis cs:-15,-2) rectangle (axis cs:62,49);
\end{axis}

\end{tikzpicture}%

%% file: tikz/methods.tikz
\begin{tikzpicture}
\begin{axis}[
width=\linewidth,
height=.6\linewidth,
grid style={dotted,anthrazit},
xmajorgrids,
yminorticks=true,
ymajorgrids,
legend columns=1,
legend pos=south west,   
legend cell align={left},
legend style={fill,fill opacity=0.8},
xlabel={$E_\mathrm{b}/N_0$ [dB]},
ylabel={BLER},
legend image post style={mark indices={}},
ymode=log,
mark size=1.5pt,
xmin=1,
xmax=4.5,
ymin=1.5e-04,
ymax=5e-01
]

\addplot[color=apfelgruen,line width = 1pt, solid,mark size=2.5pt, mark options={solid}]
table[col sep=comma]{
1.00, 4.768e-01
1.50, 2.816e-01
2.00, 1.396e-01
2.50, 5.491e-02
3.00, 1.766e-02
3.50, 4.574e-03
4.00, 9.390e-04
4.50, 1.646e-04
};
\label{plot:5g132_bp}
\addlegendentry{\footnotesize BP, conventional};

\addplot[color=lila,line width = 1pt, solid,mark=square,mark size=2.5pt, mark options={solid}, mark repeat=3, mark phase=1]
table[col sep=comma]{
1.00, 4.627e-01
1.50, 2.862e-01
2.00, 1.421e-01
2.50, 5.842e-02
3.00, 1.903e-02
3.50, 4.975e-03
4.00, 1.021e-03
4.50, 1.983e-04
};
\label{plot:5g132_bp_row_ops}
\addlegendentry{\footnotesize BP, row operations};

\addplot[color=lila,line width = 1pt, dashed,mark=square,mark size=2.5pt, mark options={solid}, mark repeat = 3, mark phase = 1 ]
table[col sep=comma]{
1.00, 3.935e-01
1.50, 2.418e-01
2.00, 1.028e-01
2.50, 3.306e-02
3.00, 8.399e-03
3.50, 1.744e-03
4.00, 2.347e-04
};
\label{plot:5g132_aed_row_ops}
\addlegendentry{\footnotesize AED-11, row operations};

\addplot[color=anthrazit,line width = 1pt, solid,mark=o,mark size=2.5pt, mark options={solid}, mark repeat = 3, mark phase = 2 ]
table[col sep=comma]{
1.00, 4.579e-01
1.50, 2.789e-01
2.00, 1.382e-01
2.50, 5.625e-02
3.00, 1.820e-02
3.50, 4.615e-03
4.00, 1.082e-03
4.50, 1.808e-04
};
\label{plot:5g132_bp_overcomplete}
\addlegendentry{\footnotesize BP, overcomplete};

\addplot[color=anthrazit,line width = 1pt, dashed,mark=o,mark size=2.5pt, mark options={solid}, mark repeat = 3, mark phase = 2]
table[col sep=comma ]{
1.00, 4.188e-01
1.50, 2.360e-01
2.00, 1.077e-01
2.50, 3.363e-02
3.00, 8.765e-03
3.50, 1.631e-03
4.00, 2.418e-04
};
\label{plot:5g132_aed_overcomplete}
\addlegendentry{\footnotesize AED-11, overcomplete};

\addplot[color=mittelblau,line width = 1pt, solid,mark=x,mark size=2.5pt, mark options={solid}, mark repeat = 3, mark phase = 3]
table[col sep=comma ]{
1.00, 4.880e-01
1.50, 2.914e-01
2.00, 1.463e-01
2.50, 5.754e-02
3.00, 1.909e-02
3.50, 5.156e-03
4.00, 9.975e-04
4.50, 1.795e-04
};
\label{plot:5g132_bp_undercomplete}
\addlegendentry{\footnotesize BP, undercomplete};

\addplot[color=mittelblau,line width = 1pt, dashed,mark=x,mark size=2.5pt, mark options={solid}, mark repeat = 3, mark phase = 3 ]
table[col sep=comma]{
1.00, 3.935e-01
1.50, 2.418e-01
2.00, 1.028e-01
2.50, 3.306e-02
3.00, 8.399e-03
3.50, 1.744e-03
4.00, 2.347e-04
};
\label{plot:5g132_aed_undercomplete}
\addlegendentry{\footnotesize AED-11, undercomplete};

\addplot[color=rot,line width = 1pt, solid,mark=x, mark size=2.5pt, mark options={solid}]
table[col sep=comma]{
1.00, 1.120e-01
1.50, 3.609e-02
2.00, 9.891e-03
2.50, 1.623e-03
3.00, 2.514e-04
};
\label{plot:5g132_osd}
\addlegendentry{\footnotesize OSD-4};

\end{axis}

\end{tikzpicture}

%% file: tikz/wifi_results.tikz
\begin{tikzpicture}
\begin{axis}[
width=\linewidth,
height=.6\linewidth,
grid style={dotted,anthrazit},
xmajorgrids,
yminorticks=true,
ymajorgrids,
legend columns=1,
legend pos=south west,   
legend cell align={left},
legend style={fill,fill opacity=0.8},
xlabel={$E_\mathrm{b}/N_0$ [dB]},
ylabel={BLER},
legend image post style={mark indices={}},
ymode=log,
mark size=1.5pt,
xmin=3,
xmax=5,
ytick={1e-1, 1e-2, 1e-3,1e-4, 1e-5,1e-6},
ymin=3e-07,
ymax=6e-01
]

\addplot[color=apfelgruen,line width = 1pt, solid,mark=x,mark size=2.5pt, mark options={solid}]
table[col sep=comma]{
3.00, 4.423e-01
3.50, 9.797e-02
4.00, 7.289e-03
4.50, 3.083e-04
5.00, 1.255e-05
};
\label{plot:wifi_bp}
\addlegendentry{\footnotesize BP};

\addplot[color=apfelgruen,line width = 1pt, solid,mark=diamond*,mark size=2.5pt, mark options={solid}]
table[col sep=comma]{
3.00, 4.216e-01
3.50, 9.693e-02
4.00, 6.413e-03
4.50, 2.799e-04
5.00, 8.157e-06
};
\label{plot:wifi_bp}
\addlegendentry{\new{\footnotesize layered BP}};

\addplot[color=gray,line width = 1pt, dashed,mark=x,mark size=2.5pt, mark options={solid}]
table[col sep=comma]{
3.00, 3.905e-01
3.50, 8.197e-02
4.00, 3.890e-03
4.50, 1.029e-04
5.00, 2.284e-06
};
\label{plot:wifi_aed_undercomplete}
\addlegendentry{\footnotesize AED-12};
\addplot[color=gray,line width = 1pt, dashed,mark=diamond*,mark size=2.5pt, mark options={solid}]
table[col sep=comma]{
3.00, 3.805e-01
3.50, 7.401e-02
4.00, 2.971e-03
4.50, 5.759e-05
5.00, 7.487e-07
};
\label{plot:wifi_aed_undercomplete}
\addlegendentry{\new{\footnotesize layered-AED-12}};

\addplot[color=mittelblau,line width = 1pt, dashed,mark=x,mark size=2.5pt, mark options={solid}]
table[col sep=comma]{
3.00, 3.695e-01
3.50, 7.541e-02
4.00, 3.291e-03
4.50, 7.536e-05
5.00, 1.356e-06
};
\label{plot:wifi_aed_undercomplete}
\addlegendentry{\footnotesize AED-27};

\addplot[color=mittelblau,line width = 1pt, dashed,mark=diamond*,mark size=2.5pt, mark options={solid}]
table[col sep=comma]{
3.00, 3.595e-01
3.50, 6.765e-02
4.00, 2.584e-03
4.50, 3.878e-05
5.00, 3.753e-07
};
\label{plot:wifi_aed_undercomplete}
\addlegendentry{\new{\footnotesize layered-AED-27}};

\addplot[color=rot,line width = 1pt,solid,mark size=2.5pt, mark options={solid}]
table[col sep=comma]{
3.00, 4.840e-02
3.25, 8.287e-03
3.50, 1.325e-03
3.75, 1.355e-04
4.00, 9.600e-06
};
\label{plot:wifi_ml}
\addlegendentry{\footnotesize ML \cite{kldatabase}};

\end{axis}

\end{tikzpicture}

%% file: tikz/5g_132_results.tikz
\begin{tikzpicture}
\begin{axis}[
width=\linewidth,
height=.57\linewidth,
grid style={dotted,anthrazit},
xmajorgrids,
yminorticks=true,
ymajorgrids,
legend columns=1,
legend pos=south west,   
legend cell align={left},
legend style={fill,fill opacity=0.8},
xlabel={$E_\mathrm{b}/N_0$ [dB]},
ylabel={BLER},
legend image post style={mark indices={}},
ymode=log,
mark size=1.5pt,
xmin=1,
xmax=4.5,
ymin=1.5e-04,
ymax=5e-01
]

\addplot[color=apfelgruen,line width = 1pt, solid,mark=x,mark size=2.5pt, mark options={solid}]
table[col sep=comma]{
1.00, 4.768e-01
1.50, 2.816e-01
2.00, 1.396e-01
2.50, 5.491e-02
3.00, 1.766e-02
3.50, 4.574e-03
4.00, 9.390e-04
4.50, 1.646e-04
};
\label{plot:5g132_bp}
\addlegendentry{\footnotesize $N=132$, BP};

\addplot[color=mittelblau,line width = 1pt, dashed,mark=x,mark size=2.5pt, mark options={solid}]
table[col sep=comma]{
1.00, 3.935e-01
1.50, 2.418e-01
2.00, 1.028e-01
2.50, 3.306e-02
3.00, 8.399e-03
3.50, 1.744e-03
4.00, 2.347e-04
};
\label{plot:5g132_aed_undercomplete}
\addlegendentry{\footnotesize $N=132$, AED-11};

\addplot[color=rot,line width = 1pt, solid,mark=x, mark size=2.5pt, mark options={solid}]
table[col sep=comma]{
1.00, 1.120e-01
1.50, 3.609e-02
2.00, 9.891e-03
2.50, 1.623e-03
3.00, 2.514e-04
};
\label{plot:5g132_osd}
\addlegendentry{\footnotesize $N=132$, OSD-4};

\addplot[color=apfelgruen,line width = 1pt, solid,mark=square,mark size=2.5pt, mark options={solid}]
table[col sep=comma]{
1.00, 4.404e-01
1.50, 1.929e-01
2.00, 5.663e-02
2.50, 1.255e-02
3.00, 1.896e-03
3.50, 1.873e-04
};
\label{plot:5g264_bp}
\addlegendentry{\footnotesize $N=264$, BP};

\addplot[color=mittelblau,line width = 1pt, dashed,mark=square,mark size=2.5pt, mark options={solid}]
table[col sep=comma]{
1.00, 3.643e-01
1.50, 1.523e-01
2.00, 3.506e-02
2.50, 6.526e-03
3.00, 5.062e-04
3.50, 2.998e-05
};
\label{plot:5g264_aed}
\addlegendentry{\footnotesize $N=264$, AED-22};

\end{axis}

\end{tikzpicture}

%% file: tikz/ccsds_results.tikz
\begin{tikzpicture}
\begin{axis}[
width=\linewidth,
height=.57\linewidth,
grid style={dotted,anthrazit},
xmajorgrids,
yminorticks=true,
ymajorgrids,
legend columns=1,
legend pos=south west,   
legend cell align={left},
legend style={fill,fill opacity=0.8},
xlabel={$E_\mathrm{b}/N_0$ [dB]},
ylabel={BLER},
legend image post style={mark indices={}},
ymode=log,
mark size=1.5pt,
xmin=1,
xmax=5,
ytick={1e-1, 1e-2, 1e-3,1e-4, 1e-5},
ymin=1e-05,
ymax=8e-01
]

\addplot[color=apfelgruen,line width = 1pt, solid,mark=x,mark size=2.5pt, mark options={solid}]
table[col sep=comma]{
0.00, 9.722e-01
0.50, 8.755e-01
1.00, 7.411e-01
1.50, 5.679e-01
2.00, 3.462e-01
2.50, 1.640e-01
3.00, 6.611e-02
3.50, 1.771e-02
4.00, 3.862e-03
4.50, 6.516e-04
5.00, 8.802e-05
};
\label{plot:ccsds128_bp}
\addlegendentry{\footnotesize $N=128$, BP};

\addplot[color=mittelblau,line width = 1pt, dashed,mark=x,mark size=2.5pt, mark options={solid}]
table[col sep=comma]{
1.00, 7.051e-01
1.50, 5.212e-01
2.00, 2.918e-01
2.50, 1.232e-01
3.00, 4.244e-02
3.50, 1.017e-02
4.00, 1.459e-03
4.50, 1.566e-04
5.00, 1.201e-05
};
\label{plot:ccsds128_aed_undercomplete}
\addlegendentry{\footnotesize $N=128$, AED-16};

\addplot[color=magenta,line width = 1pt, dotted,mark=x,mark size=2.5pt, mark options={solid}]
table[col sep=comma]{
1.00, 6.088e-01
1.50, 4.167e-01
2.00, 1.952e-01
2.50, 7.876e-02
3.00, 2.464e-02
3.50, 6.446e-03
4.00, 1.082e-03
4.50, 1.435e-04
5.00, 1.744e-05
};
\label{plot:ccsds128_sbp}
\addlegendentry{\footnotesize \new{$N=128$, SBP-16}};

\addplot[color=rot,line width = 1pt, solid,mark=x,mark size=2.5pt, mark options={solid}]
table[col sep=comma]{
1.00, 1.064e-01
1.50, 3.397e-02
2.00, 8.773e-03
2.50, 1.168e-03
3.00, 1.321e-04
3.50, 1.022e-05
};
\label{plot:ccsds128_ml}
\addlegendentry{\footnotesize $N=128$, ML \cite{kldatabase}};

\addplot[color=apfelgruen,line width = 1pt, solid,mark=square,mark size=2.5pt, mark options={solid}]
table[col sep=comma]{
0.00, 9.906e-01
0.50, 9.545e-01
1.00, 8.394e-01
1.50, 5.826e-01
2.00, 2.994e-01
2.50, 8.671e-02
3.00, 2.081e-02
3.50, 3.333e-03
4.00, 2.909e-04
4.50, 2.007e-05
};
\label{plot:ccsds256_bp}
\addlegendentry{\footnotesize $N=256$, BP};

\addplot[color=magenta,line width = 1pt, dotted,mark=square,mark size=2.5pt, mark options={solid}]
table[col sep=comma]{
1.00, 7.509e-01
1.50, 4.271e-01
2.00, 1.676e-01
2.50, 4.871e-02
3.00, 8.384e-03
3.50, 1.018e-03
4.00, 8.084e-05
};
\label{plot:ccsds256_sbp}
\addlegendentry{\footnotesize \new{$N=256$, SBP-32}};

\addplot[color=mittelblau,line width = 1pt, dashed,mark=square,mark size=2.5pt, mark options={solid}]
table[col sep=comma]{
1.00, 8.203e-01
1.50, 5.111e-01
2.00, 2.153e-01
2.50, 5.811e-02
3.00, 9.559e-03
3.50, 8.103e-04
4.00, 4.165e-05
};
\label{plot:ccsds256_aed_undercomplete}
\addlegendentry{\footnotesize $N=256$, AED-32};

\end{axis}

\end{tikzpicture}